\documentclass[oneside,12pt]{amsart}
\usepackage{amssymb,amscd,amsmath,latexsym}
\usepackage{amsthm}
\usepackage{layout}
\usepackage[mathcal]{euscript}
\usepackage{graphicx}
\usepackage[all]{xy}
\CompileMatrices

\def\U{{\bf U}}

\begin{document}

\title {The Topology of Grover Algorithm}

\author[Ali Nabi Duman]{Ali Nabi Duman}
\address{Department of Mathematics,
College of Science, University of Bahrain, Bahrain}
\email{aduman@uob.edu.bh}

\date{\today}

\maketitle
\def\SS{{\mathbb S}}
\def\G{{\mathcal G}}
\def\B{{\mathcal B}}
\def\H{{\mathcal H}}
\def\N{{\mathcal N}}
\def\K{{\mathcal K}}
\def \x{{\bf x}}
\def \M{{\mathcal M}}
\def \C{{\mathbb C}}
\def\HH{{\mathbb H}}
\def \Z{{\mathbb Z}}
\def \R{{\mathbb R}}
\def \Q{{\mathbb Q}}
\def \U{{\mathcal U}}
\def \E{{\mathcal E}}
\def \z{{\bf z}}
\def \m{{\bf m}}
\def \k{{\bf k}}
\def \n{{\bf n}}
\def \g{{\bf g}}
\def \h{{\bf h}}
\def \V{{\mathcal V}}
\def \W{{\mathcal W}}
\def \T{{\mathbb T}}
\def \X{{\mathcal X}}
\def \Y{{\mathcal Y}}
\def \P{{\bf P}}
\def \F{{\bf F}}
\def \p{{\mathfrak p}}
\def \LL{{\mathfrak L}}
\def \L{{\mathcal L}}
\def \O{{\mathfrak O}}
\def \longto{\longrightarrow}
\def \sl{{\frak sl}}
\def \C{{\mathbb C}}
\def \Q{{\mathbb Q}}
\def \vCech{{\v{C}ech\ }}
\def \calP{{\mathcal P}}
\def \H{{\mathcal H}}
\def \Cbar{{\bar{\mathcal C}}}
\def \N{{\mathbb N}}
\def \S{{\Sigma}}
\def \s{{\sigma}}
\def \T{{\mathbb T}}
\def \t{{\tau}}
\def \z{{\bar z}}
\def \bZ{{\bar Z}}
\def \P{{\mathbb P}}
\def \R{{\mathbb R}}
\def \HH{{\mathbb {H}}}
\def \div{{\rm div}}
\def \W{{\mathbb W}}
\def \L{{\mathcal L}}
\def \l{{\lambda}} \def \bl{{\Lambda}}
\def \bu{\bullet}
\def \fd{{\bullet}}
\def \Z{{\mathbb Z}}
\def \e{{\varepsilon}}
\def \ag{{\frak{g}}}
\def \ah{{\frak{h}}}
\def \O{{\mathcal O}}
\def \M{{\mathcal M}}
\def \K{{\mathcal K }}
\def \CC{{\mathcal C}}
\def \XC{{{X}_{\mathcal C}}}
\def \MC{{{\mathcal M}_{\mathcal C}}}
\def \cf {{\mathcal F}}
\def \w{{\wedge}}
\def \x{{\bf x}}
\def \k{{\kappa}}
\def \XCbar{{{X}_{\bar{\mathcal C}}}}
\def \MCbar{{{\mathcal M}_{\bar{\mathcal C}}}}
\def \i{{\sqrt{-1}}}
\def \p{{\partial}}
\def \b{{\delta}}
\def \D{{\Delta}}
\def \G{{\mathcal G}}
\def \o{{\omega}}
\def \g{{\gamma}}
\def \re {\noindent {\it Remark\ \ }}
\def \proof{{\noindent{\it Proof.\ \ }}}

\newtheorem{Th}{Theorem}[section]
\newtheorem{cor}[Th]{Corollary}
\newtheorem{lem}[Th]{Lemma}
\newtheorem{prop}[Th]{Proposition}
\newtheorem{claim}[Th]{Claim}

\theoremstyle{definition}
\newtheorem{dfn}[Th]{Definition}
\newtheorem{example}[Th]{Example}
\theoremstyle{remark}
\newtheorem*{remark}{Remark}
\newtheorem{lemma}[Th]{Lemma}

\begin{abstract}

It has been shown in recent years that quantum information has a topological nature (\cite{AC}, \cite{Co}, \cite{Co2}). In \cite{V}, Vicary undergoes the
study of quantum algorithms using this new topological approach. The advantage of this new formalism is that it allows new proofs of correctness of the
algorithms such as Deutsch-Jozsa, hidden subgroup and single-shot Grover algorithms. It also provide more clear insight to the generalizations of these
algorithms. In this paper, we consider (multi-step) Grover algorithm from this new perspective.

\end{abstract}

\section{Introduction}

Recent works demonstrated that the quantum procedures can be examined more clearly using a topological approach rather than the classical presentations
which utilize the matrices of complex numbers in an orderly manner (\cite{AC}, \cite{Co}, \cite{Co2}). In this new approach pioneered by the works of
Abramsky and Coecke one replaces the matrices with geometrical primitives.

Vicary has recently provided the topological analysis of Deutsch-Jozsa, hidden subgroup and single-shot Grover algorithms by overcoming the difficulties
that come from the use of an oracle and group representation theory \cite{V}. His work not only enable new and simple proofs of correctness but also
produce new generalizations of these algorithms.

 In \cite{V}, the author focuses on the case where a single iteration of the Grover algorithm is sufficient to find a marked element in a set. This
 algorithm can work correctly if exactly $\frac{1}{4}$ of the elements are marked. In this paper, we consider in a similar fashion the more general case where
 one element is marked in a set $S$ of size $2^n$. We perform $\sqrt{|S|}$ iteration using topological formalism.

 This paper is organized as follows: In section 2, we explain conventional Grover algorithm. Section 3 is devoted to the topological semantics that we use
 in the paper. Finally, we give the presentation of the Grover algorithm using topological semantics and prove its correctness in section 4.

 \section{Conventional Grover Search Algorithm}
Grover search algorithm is introduced to find a marked element in an array $S=\{0,1\}^n$ of size $|S|=2^n$ (see \cite{G}). The classical algorithms do this
job in $|S|/2$ queries in average while it takes $\sqrt{|S|}$ queries for Grover search algorithm. The marked element $x_0$ is defined using an indicator
function $f:S\rightarrow \mathbb{Z}_2$ where $\mathbb{Z}_2$ is the group of integers under addition modulo $2$. Here $f(x)$ is equal to $1$ for $x=x_0$ and
to $0$ for all other elements of $S$.

We then define the unitary operator $U_f:(\mathbb{C}^2)^n \otimes \mathbb{C}^2\rightarrow (\mathbb{C}^2)^n \otimes \mathbb{C}^2$ as follows:
$$|x\rangle \otimes |y\rangle \mapsto |x\rangle \otimes (|x\rangle \oplus |f(x)\rangle).$$
Here $x$ is an element of $S$ and $y$ is an element of $\mathbb{Z}_2$. The symbol $\oplus$ is the addition in the group $\mathbb{Z}_2$.

We can now introduce the quantum circuit for the Grover algorithm as follows:
\begin{center}
\includegraphics[scale=0.7]{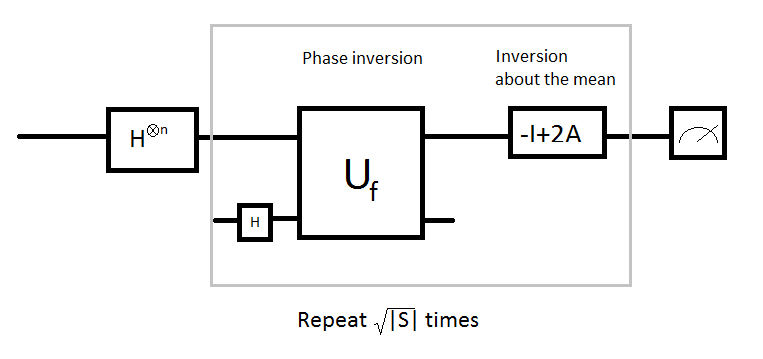}
\end{center}
where $H^{\otimes n}$ is $n$-fold Hadamard gate and the unitary matrix $-I+2A$ is the inversion about the mean. $U_f$ and $-I+2A$ combined together
separates the desired state after $\sqrt{|S|}$ consecutive applications. In other words, after $\sqrt{|S|}$th step the probability of measuring any state
except $x_0$ will be close to zero while probability of measuring $x_0$ will be greater than $1/2$. Hence, the final measurement in the algorithm will give
us the desired element $x_0$.

\section{Topological Semantics}

In this section we present topological semantics that we need to build the Grover algorithm. These geometric primitives correspond to properties of finite
groups and finite sets. The content of this section can be found in the appendix of \cite{V}. One can also refer to \cite{KL} and \cite{S} for the
mathematical foundation of the notation.

Here we start with the identity map on a finite Hilbert spaces. This is represented by a vertical wire.
\begin{center}
\includegraphics[scale=0.3]{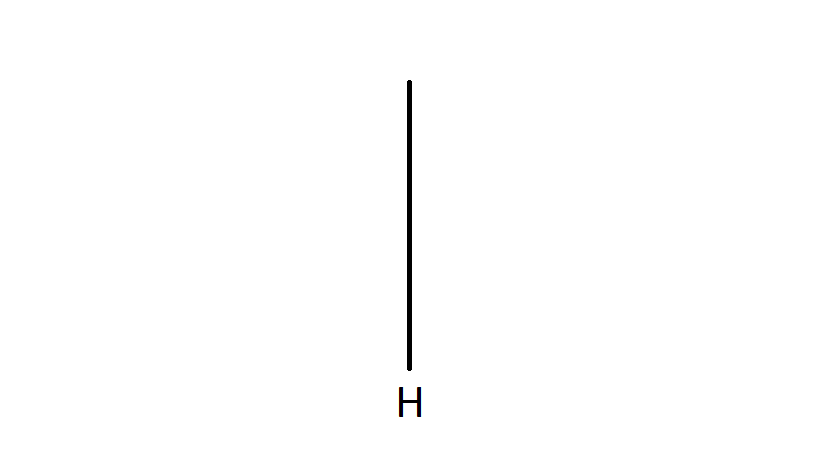}
\end{center}

 The following diagram
represent a linear map $p:H \rightarrow J.$
\begin{center}
\includegraphics[scale=0.3]{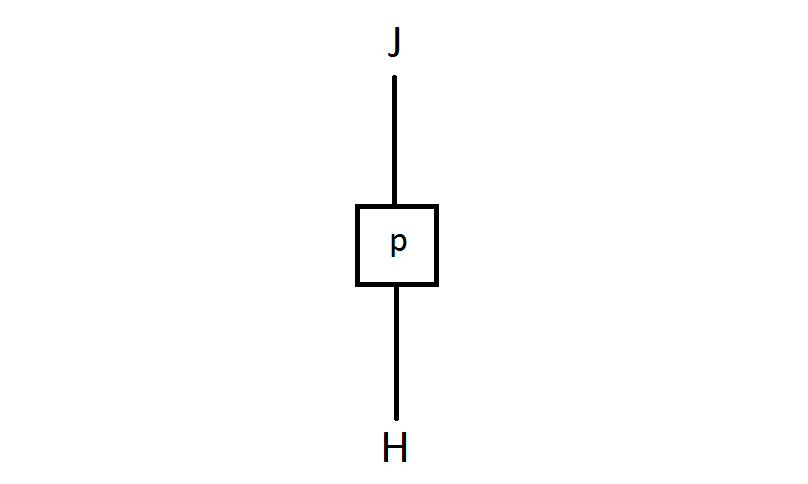}
\end{center}

 The identity on the $1$-dimensional Hilbert space is represented as the empty diagram: \vspace{2cm} \newline
One can change the relative heights of the boxes and move the components around.
\begin{center}
\includegraphics[scale=0.5]{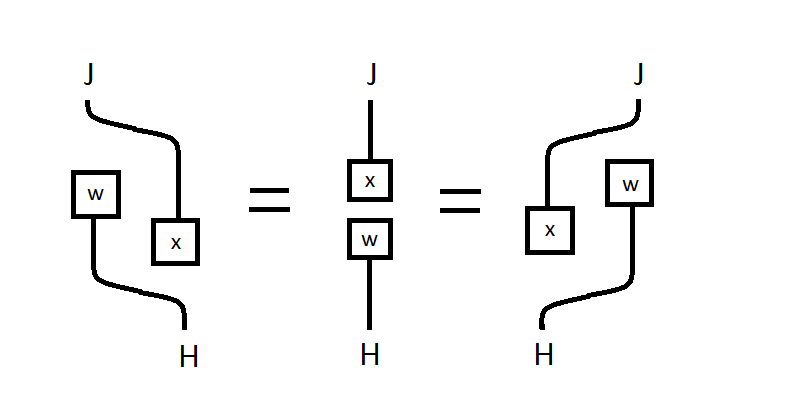}
\end{center}
Let $\mathbb{C}[S]$ be the complex vector space whose basis elements are the elements of a finite set $S$. For the sake of brevity, we denote this vector
space by $S$. Let $|s\rangle$ be a basis element of $S$. We define the maps $m:S\otimes S \rightarrow S$ and $u: \mathbb{C}\rightarrow S$ as follows:
$$m(|i\rangle \otimes |j\rangle):= \delta_{i,j}|i|rangle$$ $$u:=\sum_{i} |i\rangle.$$ The maps $m$ and $u$ gives $S$ a commutative algebra structure and
they are represented as follows:
\begin{center}
\includegraphics[scale=0.5]{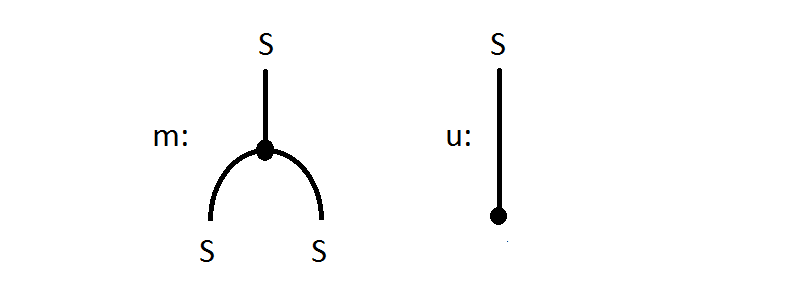}
\end{center}
 Their adjoints $m^{\dag}$ and $u^{\dag}$ are constructed in the following way:
 \begin{center}
\includegraphics[scale=0.5]{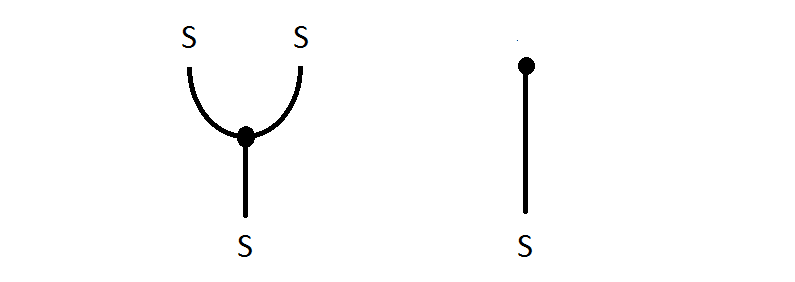}
\end{center}
A function $f:S \rightarrow T$ between finite sets extends to a linear map by $f|s\rangle := |f(s) \rangle$. The following diagram corresponds to this
function:
 \begin{center}
\includegraphics[scale=0.3]{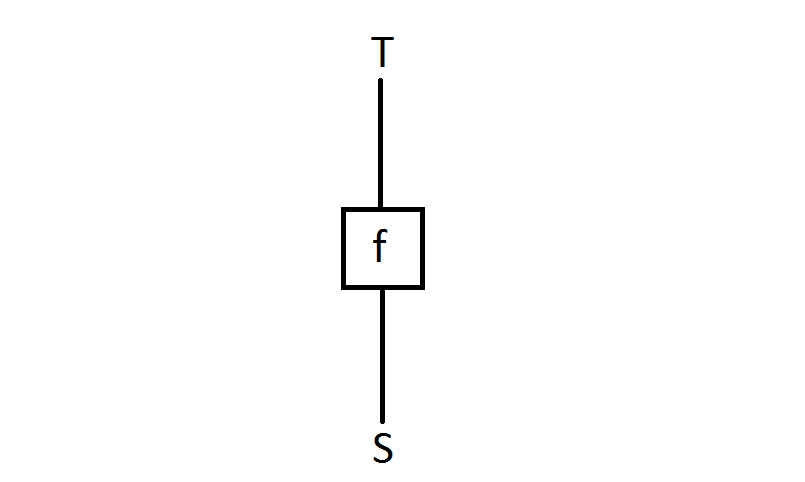}
\end{center} We have also morphisms with following two conditions  \begin{center}
\includegraphics[scale=0.5]{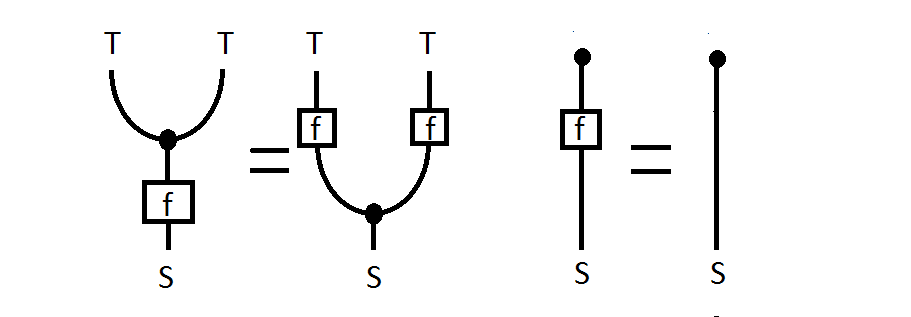}
\end{center} These are called \emph{comonoid homomorphism} conditions.

A chosen element $x\in S$ corresponds to a function $x:1\rightarrow S$, which we denote graphically as: \begin{center}
\includegraphics[scale=0.5]{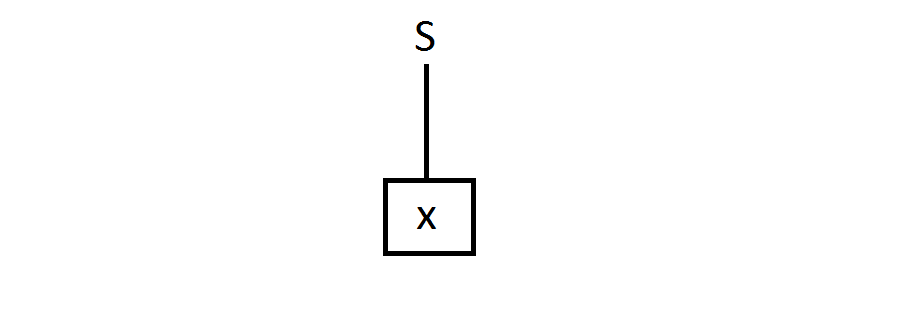}
\end{center} The deletion operator
$u^{\dag}$ reduce the elements to unit scalars:
\begin{center}
\includegraphics[scale=0.5]{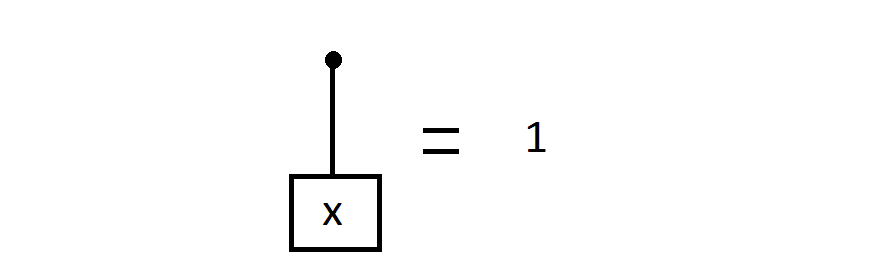}
\end{center}
$x$ and its adjoint $x^{\dag}$ are related as follows:
\begin{center}
\includegraphics[scale=0.5]{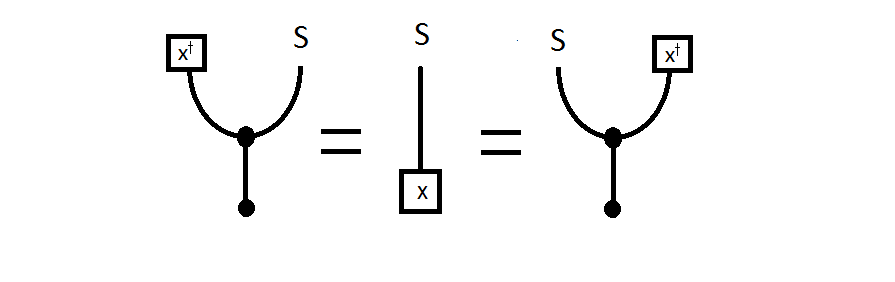}
\end{center}
Let $G$ be a group. Linearizing the multiplication function $m:G\times G\rightarrow G$ and unit element $e\in G$, we obtain the following diagrams
\begin{center}
\includegraphics[scale=0.5]{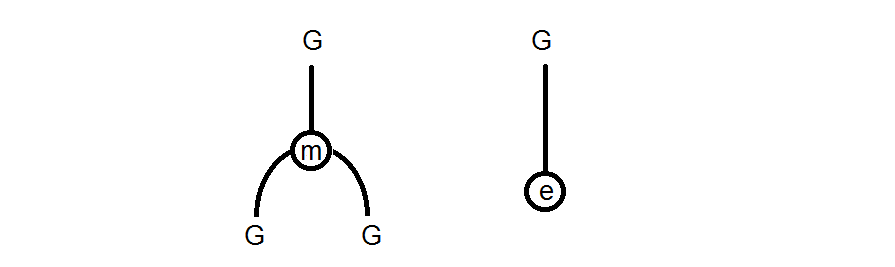}
\end{center}
 For finite group $G$, the representations are defined by the maps $\rho: G\rightarrow Mat(n)$ where $Mat(n)$ is the algebra of $n\times n$
matrices. The representations have the following properties:
\begin{center}
\includegraphics[scale=0.5]{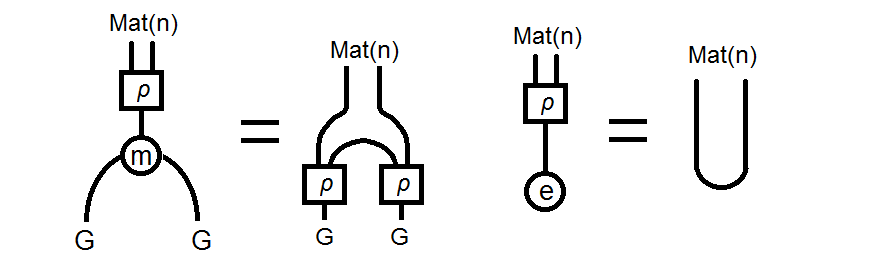}
\end{center}
We finalize this section with the following equation which is the result of basic theorems in representation theory (the reader can refer to \cite{V} for
detailed explanation):
\begin{center}
\includegraphics[scale=0.5]{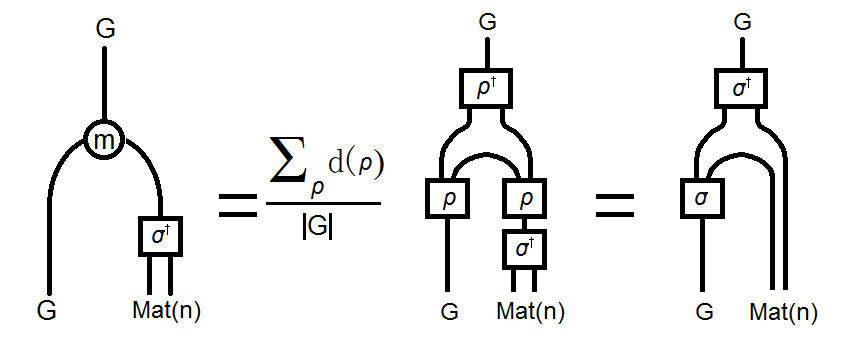}
\end{center}
 Here the sum is over equivalence classes of irreducible representations of $G$ and $\sigma^{\dag}$ is adjoint of an irreducible
representation.
\section{Topology of Grover Algorithm}
The topological structure of the (multi-step) Grover algorithm as follows:
\begin{center}
\includegraphics[scale=0.5]{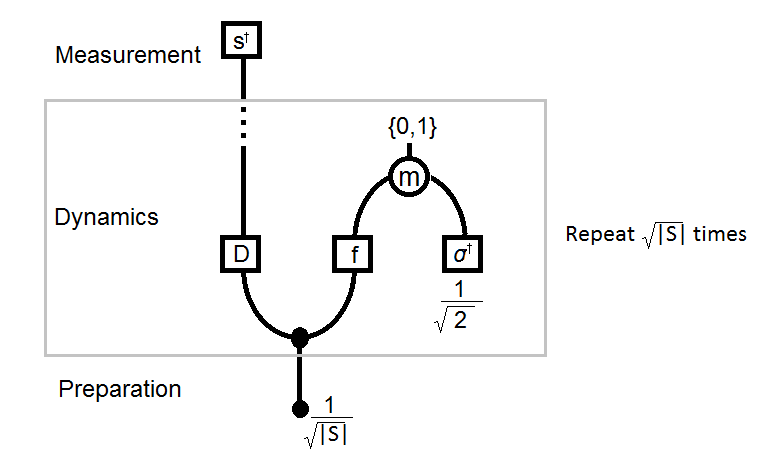}
\end{center}
 The projective measurement is performed in the base system $|s \rangle$. $D$
corresponding the inversion about the mean (i.e. $-I+2A$) can be decomposed:
\begin{center}
\includegraphics[scale=0.3]{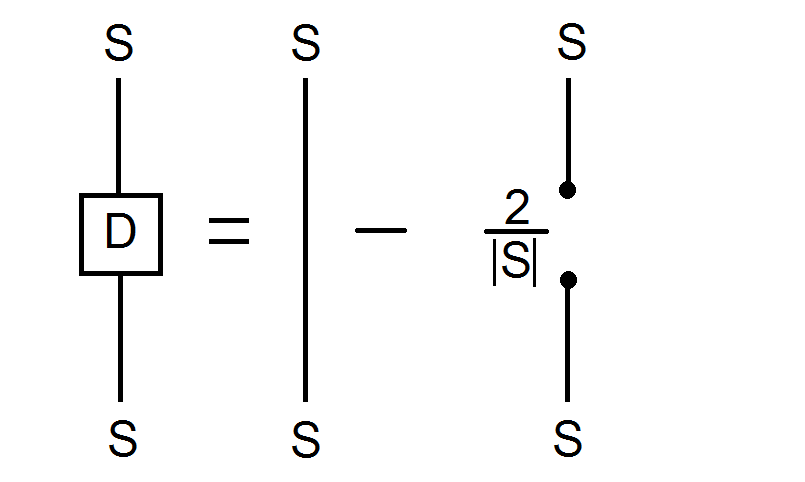}
\end{center}
 Next we use the last identity of the section 3 to
convert the algorithm to the following form:
\begin{center}
\includegraphics[scale=0.5]{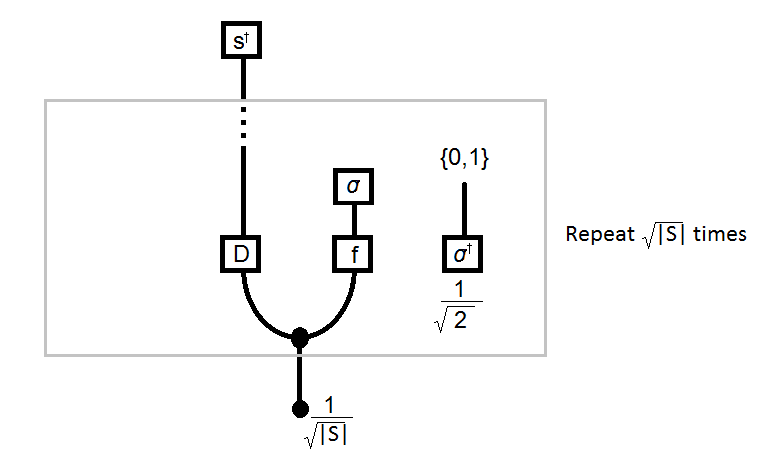}
\end{center}
 We now neglegt second part of this system as it is a product state. Using the
decomposition of $D$ we rewrite the system:
\begin{center}
\includegraphics[scale=0.5]{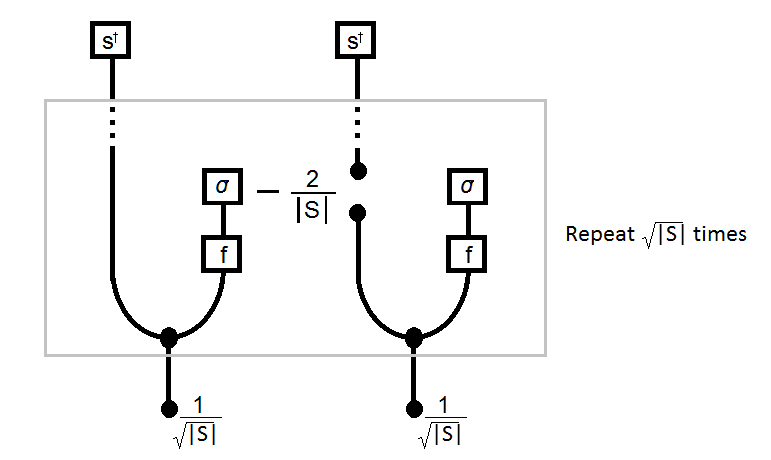}
\end{center}
\begin{center}
\includegraphics[scale=0.5]{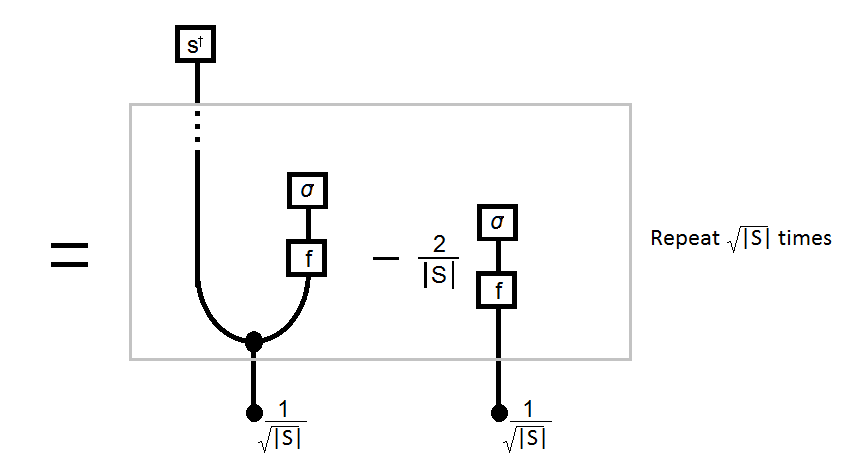}
\end{center}
 The last equality is result of the identities stated in the last section.

It is clear that for the representation $\sigma$ we have $\sum_{i\in S}\sigma=(|S|-1)+(-1)=|S|-2$ as a factor in our final measurement. For an unmarked
element $s$, we obtain the factor $+1$ in the measurement as a result of the following topological identity:
\begin{center}
\includegraphics[scale=0.5]{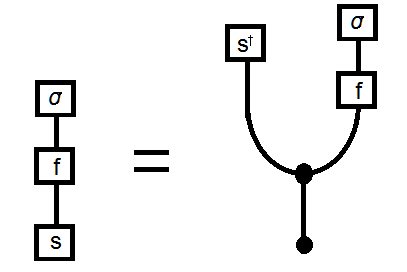}
\end{center}
Hence for such an
unmarked element we will get the following result:
$$ A=\left(\frac{1}{|S|}\right)^{\sqrt{|S|}}\left[\left(1-\frac{2}{|S|}\right)^{\sqrt{|S|}-1}\left(|S|-2\right)^{\sqrt{|S|}-1}-\frac{2}{|S|}\left(1-\frac{2}{|S|}\right)^{\sqrt{|S|}-1}\left(|S|-2\right)^{\sqrt{|S|}}\right]$$
Here the first summand in the bracket is the result of the figures which have one $s^{\dag}$ box while the second summand is from the figures without any
$s^{\dag}$ boxes. This result can be simplified to the following expression:
$$ A=\left(\frac{1}{|S|}\right)^{\sqrt{|S|}}\left(1-\frac{2}{|S|}\right)^{\sqrt{|S|}-1}\left(|S|-2\right)^{\sqrt{|S|}-1}\left(-1+\frac{4}{|S|}\right).$$
Hence $A^2$ is the probability of measuring an unmarked element. We note that $A^2$ is less than $1/2$ and approaches to zero as $n$ gets larger. The find
the probability for the marked element we now calculate the total probability for unmarked elements. Since there are $2^n-1$ unmarked elements, we consider
the value $(2^n-1)A^2$. It is straightforward to check this value is less than $1/2$ and actually goes to zero as $n$ gets larger. This implies that the
probability of measuring the marked element will be at least $1/2$ after $\sqrt{|S|}$ iterations. This proves the the correctness of the algorithm. One can
also see the reason why the algorithm is repeated $\sqrt{|S|}$ times by examining the expression $A$ where the power of the terms are the result of the the
number of the iteration.

\section{Conclusion}

In this paper, we present multi-shot Grover search algorithm using a topological semantics originated from categorical quantum mechanics program. This new
method provides a new and simple proof of the Grover algorithm and opens the ways to new generalizations.

\end{document}